\documentclass[sigconf,natbib=true,anonymous=false]{acmart}

\AtBeginDocument{%
  \providecommand\BibTeX{{%
    \normalfont B\kern-0.5em{\scshape i\kern-0.25em b}\kern-0.8em\TeX}}}

\copyrightyear{2022}
\acmYear{2022}

\usepackage{enumitem}
\usepackage{amsmath,amsfonts} 
\usepackage{algorithmic}
\usepackage{graphicx}
\usepackage{textcomp}
\usepackage{xcolor}
\def\BibTeX{{\rm B\kern-.05em{\sc i\kern-.025em b}\kern-.08em
    T\kern-.1667em\lower.7ex\hbox{E}\kern-.125emX}}
\usepackage{algorithmic}    
\usepackage{booktabs}
\usepackage[normalem]{ulem}
\useunder{\uline}{\ul}{}

\usepackage{changepage}
\usepackage{pgfplots}

\usepackage{subcaption}
\usepackage{multirow}
\usepackage{mathtools} 
\usepackage[font=normalsize]{caption}
\pgfplotsset{compat=1.16}
\usepackage{fancyhdr}

\usepackage{balance}

\let\oldmaketitle\maketitle
\renewcommand{\maketitle}{%
  \oldmaketitle%
  \thispagestyle{fancy}}
  
\begin{document}

\pagestyle{fancy}
\fancyhf{}

\cfoot{\thepage}

\title{Joint Triplet Loss Learning for \\ Next New POI Recommendation}


\author{Nicholas Lim}
\affiliation{GrabTaxi Holdings, Singapore}
\email{nic.lim@grab.com}

\author{Bryan Hooi}
\affiliation{Institute of Data Science and School of Computing, National University of Singapore, Singapore}
\email{dcsbhk@nus.edu.sg}

\author{See-Kiong Ng}
\affiliation{Institute of Data Science and School of Computing, National University of Singapore, Singapore}
\email{seekiong@nus.edu.sg}

\author{Yong Liang Goh}
\affiliation{GrabTaxi Holdings, Singapore}
\email{yongliang.goh@grab.com}



\begin{abstract}
Sparsity of the User-POI matrix is a well established problem for next POI recommendation, which hinders effective learning of user preferences. Focusing on a more granular extension of the problem, we propose a Joint Triplet Loss Learning (JTLL) module for the Next New ($N^2$) POI recommendation task, which is more challenging. Our JTLL module first computes additional training samples from the users' historical POI visit sequence, then, a designed triplet loss function is proposed to decrease and increase distances of POI and user embeddings based on their respective relations. Next, the JTLL module is jointly trained with recent approaches to additionally learn unvisited relations for the recommendation task. Experiments conducted on two known real-world LBSN datasets show that our joint training module was able to improve the performances of recent existing works.
  

\end{abstract}



\maketitle

\section{Introduction}
High sparsity of the User-POI matrix (i.e. percentage of zeroes in the matrix) is a long-standing problem that hinders effective learning of real-world predictive tasks, such as the Next New ($N^2$) POI recommendation task, a task designed to predict and recommend a set of POIs for users to visit next, where they have \emph{never} visited these POIs before \cite{hmt_grn}. While there are numerous existing works that seek to learn and alleviate sparsity of the User-POI matrix, these works include the additional incorporation of POI categories \cite{zhao2021hierarchical,categoryAware,lbpr,atca_gru,hct, zang2021cha,dong2021exploiting,lidiscovering,SNPR,chen2021multi,zhang2021interactive}, the modeling of spatio-temporal relations \cite{hstlstm}, the learning of several User-Region matrices of different levels of granularity to better learn the sparse User-POI relationships \cite{hmt_grn}, and others, proving to be effective methods for the recommendation task. However, an often overlooked relation that can also help to alleviate sparsity between users and POIs, are the POIs which the users has \emph{never} visited before, or the \emph{unvisited} relation. For example, given the search space of POIs in a region $\{l_{1}, l_{2}, l_{3}, l_{4}, l_{5}, l_{6}\}$ to determine the next POI, where a user $u_{m}$ has visited $l_{1}$ and $l_{2}$ before in her historical records, most existing works had focused on the learning of the \emph{visited} relation (i.e. $u_{m}$ visited $\{l_{1}, l_{2}\}$) for the personalized recommendation task, however, the unvisited POIs (i.e. $\{l_{3}, l_{4}, l_{5}, l_{6}\}$) are often not directly used to better learn the User-POI relations for this task. For instance, the \emph{dissimilarity} between $u_{m}$ and $\{l_{3}, l_{4}, l_{5}, l_{6}\}$ can also be used to enrich the learning of user and POI representations, while learning the similarity between $u_{m}$ and $\{l_{1}, l_{2}\}$. Further, the additional learning of this unvisited relation can help to alleviate sparsity, given the severe sparsity problem of the User-POI matrix, for example, there now exist a dissimilarity relation between user $u_{m}$ and $\{l_{3}, l_{4}, l_{5}, l_{6}\}$ that can be learned, but was not possible with only using the visited relation by most existing works. Although most existing works have learned POI-POI relations between visited POIs by the user (e.g.  $\{l_{1}, l_{2}\}$ for user $u_{m}$) and the unvisited POIs (e.g. $\{l_{3}, l_{4}, l_{5}, l_{6}\}$), such as via spatial-temporal-preference factors \cite{stp-udgat}, the unvisited relation, however, which is between the user $u_{m}$ herself, and her own unvisited POIs $\{l_{3}, l_{4}, l_{5}, l_{6}\}$, has not been learned. Motivated by this, in this paper, we propose a novel Joint Triplet Loss Learning (JTLL) module to learn both visited and unvisited relations for the $N^2$ POI recommendation task to further support sparsity alleviation. Specifically, we first design a triplet loss based loss function to reduce the distance of user and POI embeddings (user visited POI before), and increase the distance of user and POI embeddings (user never visited POI before). Then, we incorporate our JTLL module into existing works to better learn the User-POI matrix with a joint training framework.


To summarize, the following are the contributions of this paper:

\begin{itemize}[leftmargin=*,topsep=0pt]

\item We propose a novel JTLL module to further alleviate the data sparsity problem by learning both visited and unvisited relations for the $N^2$ POI recommendation task. To the best of our knowledge, this is the first work which proposes a triplet loss based loss function, that uses User-POI visited and unvisited relations for this recommendation task.

\item Our JTLL module includes a designed triplet loss function to learn both visited and unvisited relations directly. Further, we use a joint training framework with existing recommendation models to better learn User-POI relations.

\item Experiments conducted on two standard real-world LBSN datasets show that our joint training module was able to improve the performances of recent existing works.

\end{itemize}

\section{Related Work}

Extending from the next POI recommendation problem, a more challenging problem of Next New ($N^2$) POI recommendation have received recent research interest. As this $N^2$ POI recommendation task focuses on only recommending unvisited or new POIs where the user has never visited before, approaches proposed cannot merely rely on users' historical check-in sequence to perform well for this task. Early works explored conventional collaborative filtering and sequential approaches such as Matrix Factorisation (MF) and Markov Chains (MC) respectively. For example, \cite{FPMC-LR} extended the Factorizing Personalized Markov Chain (FPMC) approach \cite{FPMC} that integrates both MF and MC, to include localised region constraints and recommend nearby POIs for the $N^2$ POI recommendation task. PRME-G \cite{PRME}, a metric
embedding method, models both POIs and users in a sequential transition space and a user preference space respectively. As their method is not based on FPMC, it avoids the drawback of the independence assumption of FPMC to model the transitions \cite{PRME}. Also a metric embedding approach, \cite{HME} jointly learns the different relationships of POI sequential transitions (POI-POI), user preferences (POI-User), regional (POI-Region) and categorical (POI-Category) information in a unified way, by projecting them on a shared low-dimensional hyperbolic space. The learned hyperbolic embeddings are used with the Einstein midpoint aggregation \cite{gulcehre2018hyperbolic,ungar2005analytic} to integrate the effect of user preferences and sequential transitions for prediction. In \cite{GLR}, they proposed GLR\_GT\_LSTM, that uses a LSTM to model users' transition behaviors with latent vectors of POIs and regions, based on temporal user preference and temporal successive transition influence, as well as the spatial influence of POIs for the $N^2$ POI recommendation task. While these existing works \cite{PRME,HME,FPMC-LR,GLR} have demonstrated effectiveness for the $N^2$ POI recommendation task, they have a limitation of only considering POI samples visited within the next 6 hour threshold of the preceding POI check-in, learning and evaluating only short term preferences of the proposed methods. Following \cite{hmt_grn}, in this work, we overcome this limitation by removing the threshold, to evaluate our proposed approaches for both short and long term preferences.

The closest related work to our JTLL module is the Personalized Ranking Metric Embedding (PRME) \cite{PRME} approach, where it jointly learns POI embeddings in the sequential transition latent space, and both User and POI embeddings in the user preference latent space. For the learning of POI embeddings, the approach uses both observed POI (i.e. visited relation), and unobserved POI (i.e. unvisited relation) to optimize the learning of POI embeddings, where the euclidean distance between the previous and next visited POI should be small, and the distance between the previous and unvisited POI should be large. Although both PRME and our JTLL approach uses both visited and unvisited relations of user visits to learn our embeddings, there are several key differences. First, our proposed loss function is triplet loss based, classically involving the three roles of an anchor, a positive, and a negative. In PRME, for both POI and user representation learning, it is instead achieved with a standard pairwise metric embedding approach (i.e. only pairs of latent vectors are used at each time for the respective latent space). Second, the visited and unvisited relations of user visits is used in their sequential transition latent space to only learn POI embeddings, whereas our loss function in Eq. \eqref{jpul:eq:1} uses the visited and unvisited relations to learn both user and POI embeddings, focusing more on the user representational learning. Third, our JTLL module is designed to overcome the limitation of existing works to additionally learn unvisited relations in a joint training framework and support sparsity alleviation, with evaluation results surpassing the state-of-the-art methods. PRME is a standalone approach that has been shown by various works \cite{stgcn,hstlstm} to perform significantly poorer than a classical LSTM model for all metrics and all datasets, whereas our best performing JTLL variant, surpassed the same LSTM baseline significantly, as shown in Table. \ref{jpul:table:2}.

\section{Preliminaries}
\paragraph{\textbf{Problem Formulation}} Let $U=\{u_{1},u_{2},...,u_{M}\}$ be a set of $M$ users and $L=\{l_{1},l_{2},...,l_{Q}\}$ be a set of $Q$ POIs. $S$ is the set of visit sequences for all users where $S=\{s_{u_{1}},s_{u_{2}},...,s_{u_{M}}\}$. Each user's sequence $s_{u_{m}}$ consist of sequential POI visits $s_{u_{m}}=\{(l_{t_{1}},loc_{t_{1}},time_{t_{1}}),\\(l_{t_{2}},loc_{t_{2}},time_{t_{2}}),...,(l_{t_{i}},loc_{t_{i}},time_{t_{i}})\}$, where $l_{t_{i}}$ is the POI visited on time step $t_{i}$, with its corresponding location coordinates $loc_{t_{i}}$, and $time_{t_{i}}$ as the timestamp of the visit made. As each user's sequence $s_{u_{m}}$ is partitioned into training and testing to predict future $N^2$ POIs, we denote the superscript $train$ and $test$ respectively (e.g. $s_{u_{m}}^{train}$ and $s_{u_{m}}^{test}$).

\paragraph{Problem 1 (Next New ($N^2$) POI Recommendation)} Given user $u_{m}$, from the sequential time steps of $t_{1}$ to $t_{i-1}$ as her historical POI visit sequence $s_{u_{m}}^{train}=\{(l_{t_{1}},loc_{t_{1}},time_{t_{1}}),(l_{t_{2}},loc_{t_{2}},time_{t_{2}}),...,(l_{t_{i-1}}, \\ loc_{t_{i-1}},time_{t_{i-1}})\}$, the $N^2$ POI recommendation task is to consider a search space of POIs from $L \setminus s_{u_{m}}^{train}$, where the historically visited POIs by the respective users are removed, to compute a $N^2$ POI ranked set $y_{t_{i}}$ for the time step $t_{i}$. Accordingly, the $N^2$ POI visited $l_{t_{i}}$, should be highly ranked within $y_{t_{i}}$.

\section{Approach}
\subsection{Joint Triplet Loss Learning (JTLL)}  
\paragraph{\textbf{Loss Function.}} Here, we propose the designed triplet loss function, based on the visited and unvisited relations of users for the recommendation task. First, we compute an additional set of training data, based on users who visit a given POI. Formally, given the available users' historically visited POIs $s_{u_{m}}^{train}$, we compute the training data $l^{train}$, which is a set of tuples $(u_{h},l_{b}) \in l^{train}$, and each tuple $(u_{h},l_{b})$ denotes the check-in event relation of user $u_{h}$ having visited POI $l_{b}$ in $s_{u_{m}}^{train}$. Next, given each training tuple $(u_{h},l_{b}) \in l^{train}$, and its embeddings of $\textbf{u}_{h}$, $\textbf{l}_{b}$ (in boldface letters) from the weight matrices of user $\textbf{W}^{User}$ and POI $\textbf{W}^{POI}$ respectively, we design the following loss function, with inspirations from the triplet loss \cite{tripletLoss} and the negative sampling loss \cite{word2vec}:
\begin{gather}
J_{loss} = -\log(\sigma(\textbf{u}_{h} \: \textbf{l}_{b}^{\top})) \: -\sum_{\textbf{u}_{n} \in l^{Neg}_{b}} \log(\sigma(-\textbf{u}_{n}  \: \textbf{l}_{b}^{\top}))
\label{jpul:eq:1}
\end{gather}
where $\sigma$ is the sigmoid activation function, POI $\textbf{l}_{b}$ serves as an \emph{anchor}, $\textbf{u}_{h}$ as a \emph{positive} user who have visited POI $l_{b}$ before,  $\textbf{u}_{n} \in l^{Neg}_{b}$ as a \emph{negative} user from the set of users $l^{Neg}_{b}$, who have \emph{never} visited POI $l_{b}$ before, and can be computed from $s_{u_{m}}^{train}$. Intuitively, optimizing the loss function reduces the distance of positive users and the anchor POI (user visited POI before), and increases the distance of negative users and the anchor POI (user never visited POI before). 

\begin{figure}[t]
  \centering
  \includegraphics[width=\linewidth-1cm,height=2.2cm]{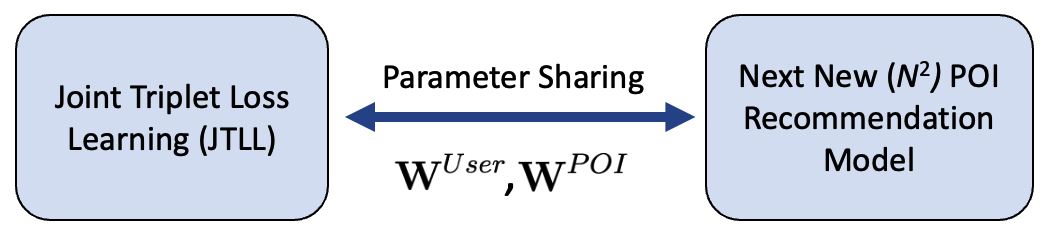}
    \caption{Proposed joint training framework.}
    \label{jpul:fig:1}
\end{figure}

\paragraph{\textbf{Joint Training.}} Our motivation of the JTLL module is to overcome the limitation of existing $N^2$ POI recommendation models to additionally learn unvisited relations. Therefore, we propose to use a joint training framework with parameter sharing, shown to be effective in other problems \cite{mao2021joint}, to easily incorporate our JTLL module and a designated existing $N^2$ POI recommendation model, where the parameters $\{\textbf{W}^{User},\textbf{W}^{POI}\}$ are shared among the two models for their own updates. Specifically, for each epoch, our JTLL module first optimizes Eq. \eqref{jpul:eq:1} with the training data $l^{train}$, with gradient steps to the user $\textbf{W}^{User}$ and POI $\textbf{W}^{POI}$ weight matrices. Then, the designated $N^2$ POI recommendation model will perform its own optimization and updates to these shared parameters, as well as other parameters unique to the model, concluding the end of a single epoch. We illustrate the proposed joint training framework in Figure \ref{jpul:fig:1}, based on our JTLL module.

\section{Experiments}

\subsection{Datasets}

We evaluate our JTLL module on two known LBSN datasets of Gowalla \cite{gowallaBrightkiteData} and Foursquare \cite{globalScaleData}. For preprocessing, we perform the same preprocessing as \cite{hmt_grn}, reproducing the same statistics in Table \ref{jpul:table:1}, where we include users with visit counts between 20 and 50 in the datasets, then removing POIs visited by less than 10 users. Accordingly, we use the first 80\% visits and the last 20\% visits of each user's sequence for training and testing respectively, based on chronological order.

\subsection{Baseline Methods and Evaluation Metrics}

\begin{itemize}[leftmargin=*,topsep=0pt]
\item \textbf{TOP}: POIs are ranked using their global frequencies in $S^{train}$ to determine popular POIs. \textbf{U-TOP} ranks POIs based on each user's historical sequence $s_{u_{m}}^{train}$, via their POI visiting frequencies.

\item \textbf{MF} \cite{MF}: A known collaborative filtering method for recommendation problems by factorizing the User-POI matrix.

\item \textbf{RNN} \cite{rnn}: A classical recurrent model that learns sequential dependencies of POI visit sequences, but suffers from vanishing gradient. The variants of \textbf{LSTM} \cite{lstm} and \textbf{GRU} \cite{gru} includes different multiplicative gates to control information flow.

\item \textbf{HST-LSTM} \cite{hstlstm}: A LSTM-based model that incorporates spatio-temporal intervals between successive POIs into the existing gates of LSTM. Following \cite{stp-udgat,stgcn}, we use their ST-LSTM variant here as the data does not include session information. \textbf{STGCN} \cite{stgcn} models short term preferences with a new cell state, as well as modeling the intervals with new distance and time gates.

\item \textbf{LSTPM} \cite{LSTPM}: A LSTM-based model that learns short term user preferences with a geo-dilated network, and long term user preferences via a nonlocal network.
\item \textbf{STAN} \cite{stan}: A bi-attention model that incorporates spatio-temporal correlations of non-contiguous visits and non-adjacent POIs.

\item \textbf{STP-UDGAT} \cite{stp-udgat}: A GAT-based method that models spatio-temporal-preference factors via different POI-POI graphs in an explore-exploit architecture.

\item \textbf{Flashback} \cite{flashback}: A RNN architecture that uses spatio-temporal intervals to compute an aggregated hidden state from past hidden states for prediction. Their best performing RNN variant is used here for evaluation. 

\item \textbf{HMT-RN} \cite{hmt_grn}: A multi-task model that learns User-POI and several User-Region matrices of different levels of granularity to alleviate sparsity. \textbf{HMT-GRN} is their best performing variant, after replacing the LSTM layer with their Graph Recurrent Network (GRN) layer.

\end{itemize}

\footnotetext[1]{https://snap.stanford.edu/data/loc-gowalla.html}
\footnotetext[2]{https://sites.google.com/site/yangdingqi/home}

\begin{table}[]
\centering
\caption{Statistics of the LBSN datasets (after preprocessing).}
\label{jpul:table:1}
\resizebox{\linewidth}{1.1cm}{%
\begin{tabular}{@{}lcccc@{}}
\toprule
Dataset & \#Country & \#User & \#POI & \#Visits \\ 
\midrule
Gowalla\textsuperscript{1} & 41 & 11,864 & 3,359 & 86,670 \\
Foursquare\textsuperscript{2} & 63 & 16,636 & 4,455 & 170,573  \\ \bottomrule
\end{tabular}%
}
\end{table}

For the recent baselines of STP-UDGAT and HMT-GRN, we use them each as the designated $N^2$ POI recommendation model in our joint learning framework (Figure \ref{jpul:fig:1}), with our JTLL module, and denoting the extended variants as \textbf{STP-UDGAT-\emph{JTLL}} and \textbf{HMT-GRN-\emph{JTLL}} accordingly. Note that for both existing works of STP-UDGAT and state-of-the-art HMT-GRN, they do not learn the unvisited relations between users and POIs in their works, therefore, following our motivation, we extend these recent works with our JTLL module to additionally learn the unvisited relations, and further alleviating sparsity.


\begin{table*}[]
\caption{Performance in $N^2$-Acc@$K$ and $N^2$-MRR for only unvisited next POI test samples.} 
\label{jpul:table:2}

\resizebox{\linewidth-2.5cm}{8.2cm}{%
\begin{tabular}{@{}lccccc@{}}
\toprule
\multicolumn{6}{c}{Gowalla} \\ \midrule
 &  $N^2$-Acc@1 & $N^2$-Acc@5 & $N^2$-Acc@10 & $N^2$-Acc@20 & $N^2$-MRR \\ \midrule
TOP & 0.0068 & 0.0281 & 0.0574 & 0.0874 & 0.0227 \\
U-TOP  & 0 & 0 & 0 & 0 & 0 \\
MF  & 0.0015$\pm$0.001 & 0.0032$\pm$0.001 & 0.0046$\pm$0.001 & 0.0073$\pm$0.001 & 0.0040$\pm$0.001 \\
RNN & 0.0356$\pm$0.001 & 0.1045$\pm$0.001 & 0.1496$\pm$0.001 & 0.2034$\pm$0.001 & 0.0746$\pm$0.001 \\
GRU  & 0.0367$\pm$0.001 & 0.1064$\pm$0.001 & 0.1563$\pm$0.001 & 0.2150$\pm$0.001 & 0.0773$\pm$0.001 \\
LSTM  & 0.0419$\pm$0.001 & 0.1140$\pm$0.001 & 0.1661$\pm$0.001 & 0.2291$\pm$0.001 & 0.0843$\pm$0.001 \\
HST-LSTM  & 0.0069$\pm$0.001 & 0.0293$\pm$0.001 & 0.0545$\pm$0.002 & 0.0854$\pm$0.001 & 0.0233$\pm$0.001 \\
STGCN  & 0.0126$\pm$0.001 & 0.0460$\pm$0.002 & 0.0777$\pm$0.003 & 0.1269$\pm$0.003 & 0.0374$\pm$0.001 \\
LSTPM  & 0.0353$\pm$0.001 & 0.0869$\pm$0.002 & 0.1199$\pm$0.002 & 0.1613$\pm$0.003 & 0.0653$\pm$0.001 \\
STAN 	& 0.0143$\pm$0.001	& 0.0513$\pm$0.002	& 0.0843$\pm$0.002	& 0.1323$\pm$0.005	& 0.0398$\pm$0.001\\
Flashback &	0.0153$\pm$0.001 &	0.0517$\pm$0.002 &	0.0825$\pm$0.003 &	0.1288$\pm$0.001 &	0.0412$\pm$0.001 \\ \midrule
STP-UDGAT & 0.0251$\pm$0.001 & 0.0712$\pm$0.001 & 0.1014$\pm$0.001 & 0.1393$\pm$0.002 & 0.0517$\pm$0.001 \\ 
STP-UDGAT-\emph{JTLL} &  0.0338$\pm$0.001	 & 0.0926$\pm$0.001 & 	0.1290$\pm$0.002 & 	0.1746$\pm$0.002 & 	0.0671$\pm$0.001
\\ \midrule
HMT-RN  & 0.0523$\pm$0.001 & 0.1334$\pm$0.001 & 0.1888$\pm$0.001 & 0.2536$\pm$0.001 & 0.0987$\pm$0.001 \\
HMT-GRN & 0.0539$\pm$0.001 & 0.1369$\pm$0.001 & 0.1920$\pm$0.001 & 0.2579$\pm$0.001 & 0.1008$\pm$0.001 \\ 
HMT-GRN-\emph{JTLL} & \textbf{0.0543$\pm$0.001} & \textbf{0.1388$\pm$0.001} & \textbf{0.1970$\pm$0.001} & \textbf{0.2637$\pm$0.001} & \textbf{0.1026$\pm$0.001} \\
				
\toprule 
\multicolumn{6}{c}{Foursquare} \\ \midrule
 & $N^2$-Acc@1 & $N^2$-Acc@5 & $N^2$-Acc@10 & $N^2$-Acc@20 & $N^2$-MRR \\ \midrule
TOP & 0.0056 & 0.0247 & 0.0373 & 0.0604 & 0.0192 \\
U-TOP  & 0 & 0 & 0 & 0 & 0 \\
MF & 0.0009$\pm$0.001 & 0.0028$\pm$0.001 & 0.0043$\pm$0.001 & 0.0061$\pm$0.001 & 0.0031$\pm$0.001 \\
RNN  & 0.0444$\pm$0.001 & 0.1272$\pm$0.001 & 0.1858$\pm$0.001 & 0.2536$\pm$0.001 & 0.0914$\pm$0.001 \\
GRU  & 0.0459$\pm$0.001 & 0.1306$\pm$0.001 & 0.1908$\pm$0.001 & 0.2644$\pm$0.001 & 0.0945$\pm$0.001 \\
LSTM & 0.0505$\pm$0.001 & 0.1400$\pm$0.001 & 0.2035$\pm$0.001 & 0.2828$\pm$0.001 & 0.1023$\pm$0.001 \\
HST-LSTM  & 0.0058$\pm$0.001 & 0.0218$\pm$0.001 & 0.0369$\pm$0.001 & 0.0591$\pm$0.001 & 0.0181$\pm$0.001 \\
STGCN & 0.0114$\pm$0.001 & 0.0497$\pm$0.002 & 0.0896$\pm$0.003 & 0.1503$\pm$0.003 & 0.0400$\pm$0.001 \\
LSTPM & 0.0426$\pm$0.001 & 0.1052$\pm$0.001 & 0.1466$\pm$0.001 & 0.1951$\pm$0.001 & 0.0782$\pm$0.001 \\
STAN  & 0.0249$\pm$0.002 & 0.0870$\pm$0.007 & 0.1384$\pm$0.008 & 0.2070$\pm$0.011 & 0.0643$\pm$0.004 \\
Flashback  &	0.0229$\pm$0.001 &	0.0742$\pm$0.001 &	0.1185$\pm$0.001 &	0.1820$\pm$0.001 &	0.0577$\pm$0.001 \\ \midrule
STP-UDGAT  & 0.0382$\pm$0.001 & 0.1156$\pm$0.002 & 0.1699$\pm$0.002 & 0.2324$\pm$0.001 & 0.0811$\pm$0.001 \\ 
STP-UDGAT-\emph{JTLL} & 0.0412$\pm$0.001	 & 0.1153$\pm$0.002 & 	0.1612$\pm$0.001	 & 0.2154$\pm$0.002 & 	0.0811$\pm$0.001
\\  \midrule
HMT-RN  & 0.0670$\pm$0.001 & 0.1738$\pm$0.001 & 0.2486$\pm$0.001 & 0.3357$\pm$0.001 & 0.1269$\pm$0.001 \\
HMT-GRN  & 0.0686$\pm$0.001 & 0.1756$\pm$0.002 & 0.2507$\pm$0.001 & 0.3386$\pm$0.001 & 0.1288$\pm$0.001 \\ 
HMT-GRN-\emph{JTLL}  & \textbf{0.0704$\pm$0.001} & \textbf{0.1779$\pm$0.001} & \textbf{0.2553$\pm$0.001} & \textbf{0.3452$\pm$0.001} & \textbf{0.1313$\pm$0.001}
\\ 
								

\bottomrule
\end{tabular}%
}
\end{table*}

\paragraph{\textbf{Metrics.}} Following \cite{hmt_grn} for the $N^2$ POI recommendation task, we use the standard metric of Acc@$K$ and Mean Reciprocal Rank (MRR), denoting them as $N^2$-Acc@$K$ for $K\in\{1,5,10,20\}$ and $N^2$-MRR respectively. Note that all test ground truth POI samples are preprocessed to include only new or unvisited POI $l_{t_{i}} \notin s_{u_{m}}^{train}$ to correctly evaluate for this recommendation task.

\subsection{Experimental Settings}
For our JTLL module, we use the Adam optimizer with a learning rate of 0.001, with a batch size of 64 tuples from the training data $l^{train}$. Further, we apply a dropout of 0.8 to the user and POI embeddings from the weight matrices $\textbf{W}^{User}$ and $\textbf{W}^{POI}$ before optimizing the loss function in Eq. \eqref{jpul:eq:1}. For all other hyperparameters (e.g. number of epochs and embedding dimension size), we set it to be the same as the designated existing $N^2$ POI recommendation model for simplicity and ease of implementation. For the baseline methods, for MF, RNN, GRU, and LSTM, we use the same settings where applicable. For the other recent works of HST-LSTM, STGCN, LSTPM, STP-UDGAT, and HMT-GRN, we follow their recommended settings as described. For Flashback and STAN, we apply grid search and use the best performing models for evaluation, as their recommended hyperparameters does not work as well in our experiments.


\subsection{Results}
We report the evaluation results of our proposed JTLL module and the baselines in Table \ref{jpul:table:2}. For all baselines and models, except TOP and U-TOP which are deterministic, we show the averaged results of 5 runs on different random seeds, as well as their respective standard deviations:

\begin{itemize}[leftmargin=*,topsep=0pt]

\item We see that the variant of HMT-GRN-\emph{JTLL} has the best performance in both Gowalla and Foursquare datasets, where existing state-of-the-art HMT-GRN is the designated $N^2$ POI recommendation model, with our JTLL module. More importantly, we can observe an unanimous increase of performance when comparing HMT-GRN-\emph{JTLL} and HMT-GRN across both datasets and all metrics, demonstrating the effectiveness of our JTLL module, as well as the necessary learning of unvisited relations for this recommendation task.

\item Comparing STP-UDGAT-\emph{JTLL} with STP-UDGAT, we similarly see a significant increase of performance for STP-UDGAT-\emph{JTLL} in the Gowalla dataset for all metrics. For the Foursquare dataset, however, we see that STP-UDGAT-\emph{JTLL} only have the best result for $N^2$-Acc@1, and not the rest of the metrics. We believe that this is due to the use of the high 0.8 dropout in our experimental setting for regularization, which indeed allowed STP-UDGAT-\emph{JTLL} to perform the best for $N^2$-Acc@1 (i.e. correctly predicting the next new POI itself), but hindered the general learning capability of the model. However, we see that the high dropout rate did not adversely affect the learning ability of the model for the Gowalla dataset, as STP-UDGAT-\emph{JTLL} always have better performances than STP-UDGAT.

\item Simpler baselines such as TOP and MF do not perform well for this task as they do not learn sequential dependencies of check-in event transitions. For U-TOP, as it ranks POIs based on historically visited POIs of the respective user, it is unable to rank new or unvisited POIs, and therefore, the scores of zeroes for all the metrics. 

\item For all other baselines, same as \cite{hmt_grn}, we believe that they do not perform as well due to the sparsity problem. As HMT-GRN is designed to alleviate sparsity, it was thus able to achieve state-of-the-art results, however, with the inclusion of our JTLL module to additionally learn unvisited relations in a novel way, it is now able to perform better.

\end{itemize}

\section{Conclusion}
This work proposed a novel JTLL module to alleviate the data sparsity problem by learning both visited and unvisited relations from users. Our JTLL module first computes the additional training data of user and POI tuples, then optimizing our designed loss function. A joint training framework is proposed to implement our JTLL module with existing $N^2$ POI recommendation models, and with parameter sharing. Experimental results on two known real-world LBSN datasets demonstrate the effectiveness of the proposed approach with notable improvements over existing works. For future work, we hope to explore other factors to further alleviate sparsity.

\section*{Acknowledgment}
This work was funded by the Grab-NUS AI Lab, a joint collaboration between GrabTaxi Holdings Pte. Ltd. and National University of Singapore, and the Industrial Postgraduate Program (Grant: S18-1198-IPP-II) funded by the Economic Development Board of Singapore.

\bibliographystyle{ACM-Reference-Format}
\bibliography{sample-base}


\begin{thebibliography}{34}


\ifx \showCODEN    \undefined \def \showCODEN     #1{\unskip}     \fi
\ifx \showDOI      \undefined \def \showDOI       #1{#1}\fi
\ifx \showISBNx    \undefined \def \showISBNx     #1{\unskip}     \fi
\ifx \showISBNxiii \undefined \def \showISBNxiii  #1{\unskip}     \fi
\ifx \showISSN     \undefined \def \showISSN      #1{\unskip}     \fi
\ifx \showLCCN     \undefined \def \showLCCN      #1{\unskip}     \fi
\ifx \shownote     \undefined \def \shownote      #1{#1}          \fi
\ifx \showarticletitle \undefined \def \showarticletitle #1{#1}   \fi
\ifx \showURL      \undefined \def \showURL       {\relax}        \fi
\providecommand\bibfield[2]{#2}
\providecommand\bibinfo[2]{#2}
\providecommand\natexlab[1]{#1}
\providecommand\showeprint[2][]{arXiv:#2}

\bibitem[\protect\citeauthoryear{Chen, Ying, Lyu, Yu, and Chen}{Chen
  et~al\mbox{.}}{2021}]%
        {chen2021multi}
\bibfield{author}{\bibinfo{person}{Ling Chen}, \bibinfo{person}{Yuankai Ying},
  \bibinfo{person}{Dandan Lyu}, \bibinfo{person}{Shanshan Yu}, {and}
  \bibinfo{person}{Gencai Chen}.} \bibinfo{year}{2021}\natexlab{}.
\newblock \showarticletitle{A multi-task embedding based personalized POI
  recommendation method}.
\newblock \bibinfo{journal}{\emph{CCF TPCI}} (\bibinfo{year}{2021}),
  \bibinfo{pages}{1--17}.
\newblock


\bibitem[\protect\citeauthoryear{{Cheng}, {Yang}, {Lyu}, and {King}}{{Cheng}
  et~al\mbox{.}}{2013}]%
        {FPMC-LR}
\bibfield{author}{\bibinfo{person}{Chen {Cheng}}, \bibinfo{person}{Haiqin
  {Yang}}, \bibinfo{person}{Michael~R. {Lyu}}, {and} \bibinfo{person}{Irwin
  {King}}.} \bibinfo{year}{2013}\natexlab{}.
\newblock \showarticletitle{Where You Like to Go Next: Successive
  Point-of-Interest Recommendation}. In \bibinfo{booktitle}{\emph{IJCAI}}.
  \bibinfo{pages}{2605--2611}.
\newblock


\bibitem[\protect\citeauthoryear{{Cho}, {Myers}, and {Leskovec}}{{Cho}
  et~al\mbox{.}}{2011}]%
        {gowallaBrightkiteData}
\bibfield{author}{\bibinfo{person}{Eunjoon {Cho}}, \bibinfo{person}{Seth~A.
  {Myers}}, {and} \bibinfo{person}{Jure {Leskovec}}.}
  \bibinfo{year}{2011}\natexlab{}.
\newblock \showarticletitle{Friendship and mobility: User movement in
  location-based social networks}. In \bibinfo{booktitle}{\emph{KDD}}.
  \bibinfo{pages}{1082–1090}.
\newblock


\bibitem[\protect\citeauthoryear{{Cho}, {Bahdanau}, {Bougares}, {Schwenk}, and
  {Bengio}}{{Cho} et~al\mbox{.}}{2014}]%
        {gru}
\bibfield{author}{\bibinfo{person}{Kyunghyun {Cho}}, \bibinfo{person}{Dzmitry
  {Bahdanau}}, \bibinfo{person}{Fethi {Bougares}}, \bibinfo{person}{Holger
  {Schwenk}}, {and} \bibinfo{person}{Yoshua {Bengio}}.}
  \bibinfo{year}{2014}\natexlab{}.
\newblock \showarticletitle{Learning Phrase Representations using RNN
  Encoder-Decoder for Statistical Machine Translation}. In
  \bibinfo{booktitle}{\emph{EMNLP}}. \bibinfo{pages}{1724--1734}.
\newblock


\bibitem[\protect\citeauthoryear{Dong, Meng, and Zhang}{Dong
  et~al\mbox{.}}{2021}]%
        {dong2021exploiting}
\bibfield{author}{\bibinfo{person}{Zheng Dong}, \bibinfo{person}{Xiangwu Meng},
  {and} \bibinfo{person}{Yujie Zhang}.} \bibinfo{year}{2021}\natexlab{}.
\newblock \showarticletitle{Exploiting Category-Level Multiple Characteristics
  for POI Recommendation}.
\newblock \bibinfo{journal}{\emph{TKDE}} (\bibinfo{year}{2021}).
\newblock


\bibitem[\protect\citeauthoryear{{Elman}}{{Elman}}{1990}]%
        {rnn}
\bibfield{author}{\bibinfo{person}{Jeffrey~L. {Elman}}.}
  \bibinfo{year}{1990}\natexlab{}.
\newblock \showarticletitle{Finding Structure in Time}. In
  \bibinfo{booktitle}{\emph{Cognitive Science 14}}. \bibinfo{pages}{179--211}.
\newblock


\bibitem[\protect\citeauthoryear{{Feng}, {Li}, {Zeng}, {Cong}, {Chee}, and
  {Yuan}}{{Feng} et~al\mbox{.}}{2015}]%
        {PRME}
\bibfield{author}{\bibinfo{person}{Shanshan {Feng}}, \bibinfo{person}{Xutao
  {Li}}, \bibinfo{person}{Yifeng {Zeng}}, \bibinfo{person}{Gao {Cong}},
  \bibinfo{person}{Yeow~Meng {Chee}}, {and} \bibinfo{person}{Quan {Yuan}}.}
  \bibinfo{year}{2015}\natexlab{}.
\newblock \showarticletitle{Personalized Ranking Metric Embedding for Next New
  POI Recommendation}. In \bibinfo{booktitle}{\emph{IJCAI}}.
  \bibinfo{pages}{2069--2075}.
\newblock


\bibitem[\protect\citeauthoryear{Feng, Tran, Cong, Chen, Li, and Li}{Feng
  et~al\mbox{.}}{2020}]%
        {HME}
\bibfield{author}{\bibinfo{person}{Shanshan Feng}, \bibinfo{person}{Lucas~Vinh
  Tran}, \bibinfo{person}{Gao Cong}, \bibinfo{person}{Lisi Chen},
  \bibinfo{person}{Jing Li}, {and} \bibinfo{person}{Fan Li}.}
  \bibinfo{year}{2020}\natexlab{}.
\newblock \showarticletitle{HME: A Hyperbolic Metric Embedding Approach for
  Next-POI Recommendation}. In \bibinfo{booktitle}{\emph{SIGIR}}.
  \bibinfo{pages}{1429--1438}.
\newblock


\bibitem[\protect\citeauthoryear{Gulcehre, Denil, Malinowski, Razavi, Pascanu,
  Hermann, Battaglia, Bapst, Raposo, Santoro, et~al\mbox{.}}{Gulcehre
  et~al\mbox{.}}{2019}]%
        {gulcehre2018hyperbolic}
\bibfield{author}{\bibinfo{person}{Caglar Gulcehre}, \bibinfo{person}{Misha
  Denil}, \bibinfo{person}{Mateusz Malinowski}, \bibinfo{person}{Ali Razavi},
  \bibinfo{person}{Razvan Pascanu}, \bibinfo{person}{Karl~Moritz Hermann},
  \bibinfo{person}{Peter Battaglia}, \bibinfo{person}{Victor Bapst},
  \bibinfo{person}{David Raposo}, \bibinfo{person}{Adam Santoro},
  {et~al\mbox{.}}} \bibinfo{year}{2019}\natexlab{}.
\newblock \showarticletitle{Hyperbolic Attention Networks}. In
  \bibinfo{booktitle}{\emph{ICLR}}.
\newblock


\bibitem[\protect\citeauthoryear{He, Li, and Liao}{He et~al\mbox{.}}{2017}]%
        {lbpr}
\bibfield{author}{\bibinfo{person}{Jing He}, \bibinfo{person}{Xin Li}, {and}
  \bibinfo{person}{Lejian Liao}.} \bibinfo{year}{2017}\natexlab{}.
\newblock \showarticletitle{Category-aware Next Point-of-Interest
  Recommendation via Listwise Bayesian Personalized Ranking}. In
  \bibinfo{booktitle}{\emph{IJCAI}}, Vol.~\bibinfo{volume}{17}.
  \bibinfo{pages}{1837--1843}.
\newblock


\bibitem[\protect\citeauthoryear{{Hochreiter} and {Schmidhuber}}{{Hochreiter}
  and {Schmidhuber}}{1997}]%
        {lstm}
\bibfield{author}{\bibinfo{person}{Sepp {Hochreiter}} {and}
  \bibinfo{person}{Jurgen {Schmidhuber}}.} \bibinfo{year}{1997}\natexlab{}.
\newblock \showarticletitle{Long Short Term Memory}. In
  \bibinfo{booktitle}{\emph{Neural Computation 9(8)}}.
  \bibinfo{pages}{1735--1780}.
\newblock


\bibitem[\protect\citeauthoryear{{Kong} and {Wu}}{{Kong} and {Wu}}{2018}]%
        {hstlstm}
\bibfield{author}{\bibinfo{person}{Dejiang {Kong}} {and} \bibinfo{person}{Fei
  {Wu}}.} \bibinfo{year}{2018}\natexlab{}.
\newblock \showarticletitle{HST-LSTM: A Hierarchical Spatial-Temporal
  Long-Short Term Memory Network for Location Prediction}. In
  \bibinfo{booktitle}{\emph{IJCAI}}. \bibinfo{pages}{2341--2347}.
\newblock


\bibitem[\protect\citeauthoryear{{Koren}, {Bell}, and {Volinsky}}{{Koren}
  et~al\mbox{.}}{2009}]%
        {MF}
\bibfield{author}{\bibinfo{person}{Y. {Koren}}, \bibinfo{person}{R. {Bell}},
  {and} \bibinfo{person}{C. {Volinsky}}.} \bibinfo{year}{2009}\natexlab{}.
\newblock \showarticletitle{Matrix Factorization Techniques for Recommender
  Systems}.
\newblock \bibinfo{journal}{\emph{Computer}} \bibinfo{volume}{42},
  \bibinfo{number}{8} (\bibinfo{year}{2009}), \bibinfo{pages}{30--37}.
\newblock


\bibitem[\protect\citeauthoryear{Li, Chen, Luo, Yin, and Huang}{Li
  et~al\mbox{.}}{2021}]%
        {lidiscovering}
\bibfield{author}{\bibinfo{person}{Yang Li}, \bibinfo{person}{Tong Chen},
  \bibinfo{person}{Yadan Luo}, \bibinfo{person}{Hongzhi Yin}, {and}
  \bibinfo{person}{Zi Huang}.} \bibinfo{year}{2021}\natexlab{}.
\newblock \showarticletitle{Discovering Collaborative Signals for Next POI
  Recommendation with Iterative Seq2Graph Augmentation}. In
  \bibinfo{booktitle}{\emph{IJCAI}}. \bibinfo{pages}{1491--1497}.
\newblock


\bibitem[\protect\citeauthoryear{Lim, Hooi, Ng, Goh, Weng, and Tan}{Lim
  et~al\mbox{.}}{2022}]%
        {hmt_grn}
\bibfield{author}{\bibinfo{person}{Nicholas Lim}, \bibinfo{person}{Bryan Hooi},
  \bibinfo{person}{See-Kiong Ng}, \bibinfo{person}{Yong~Liang Goh},
  \bibinfo{person}{Renrong Weng}, {and} \bibinfo{person}{Rui Tan}.}
  \bibinfo{year}{2022}\natexlab{}.
\newblock \showarticletitle{Hierarchical Multi-Task Graph Recurrent Network for
  Next POI Recommendation}. In \bibinfo{booktitle}{\emph{SIGIR}}.
  \bibinfo{pages}{1133–1143}.
\newblock


\bibitem[\protect\citeauthoryear{Lim, Hooi, Ng, Wang, Goh, Weng, and
  Varadarajan}{Lim et~al\mbox{.}}{2020}]%
        {stp-udgat}
\bibfield{author}{\bibinfo{person}{Nicholas Lim}, \bibinfo{person}{Bryan Hooi},
  \bibinfo{person}{See-Kiong Ng}, \bibinfo{person}{Xueou Wang},
  \bibinfo{person}{Yong~Liang Goh}, \bibinfo{person}{Renrong Weng}, {and}
  \bibinfo{person}{Jagannadan Varadarajan}.} \bibinfo{year}{2020}\natexlab{}.
\newblock \showarticletitle{STP-UDGAT: Spatial-Temporal-Preference User
  Dimensional Graph Attention Network for Next POI Recommendation}. In
  \bibinfo{booktitle}{\emph{CIKM}}. \bibinfo{pages}{845--854}.
\newblock


\bibitem[\protect\citeauthoryear{Liu, Pei, Wang, Yang, Zhang, Wang, Dai, Qi,
  and Ma}{Liu et~al\mbox{.}}{2021}]%
        {atca_gru}
\bibfield{author}{\bibinfo{person}{Yuwen Liu}, \bibinfo{person}{Aixiang Pei},
  \bibinfo{person}{Fan Wang}, \bibinfo{person}{Yihong Yang},
  \bibinfo{person}{Xuyun Zhang}, \bibinfo{person}{Hao Wang},
  \bibinfo{person}{Hongning Dai}, \bibinfo{person}{Lianyong Qi}, {and}
  \bibinfo{person}{Rui Ma}.} \bibinfo{year}{2021}\natexlab{}.
\newblock \showarticletitle{An attention-based category-aware GRU model for the
  next POI recommendation}.
\newblock \bibinfo{journal}{\emph{International Journal of Intelligent
  Systems}} (\bibinfo{year}{2021}).
\newblock


\bibitem[\protect\citeauthoryear{Lu and Huang}{Lu and Huang}{2020}]%
        {GLR}
\bibfield{author}{\bibinfo{person}{Yi-Shu Lu} {and} \bibinfo{person}{Jiun-Long
  Huang}.} \bibinfo{year}{2020}\natexlab{}.
\newblock \showarticletitle{GLR: A graph-based latent representation model for
  successive POI recommendation}.
\newblock \bibinfo{journal}{\emph{Future Generation Computer Systems}}
  \bibinfo{volume}{102} (\bibinfo{year}{2020}), \bibinfo{pages}{230--244}.
\newblock


\bibitem[\protect\citeauthoryear{Luo, Liu, and Liu}{Luo et~al\mbox{.}}{2021}]%
        {stan}
\bibfield{author}{\bibinfo{person}{Yingtao Luo}, \bibinfo{person}{Qiang Liu},
  {and} \bibinfo{person}{Zhaocheng Liu}.} \bibinfo{year}{2021}\natexlab{}.
\newblock \showarticletitle{STAN: Spatio-Temporal Attention Network for next
  Point-of-Interest Recommendation}. In \bibinfo{booktitle}{\emph{WWW}}.
  \bibinfo{pages}{2177--2185}.
\newblock


\bibitem[\protect\citeauthoryear{Mao, Shen, Yu, and Cai}{Mao
  et~al\mbox{.}}{2021}]%
        {mao2021joint}
\bibfield{author}{\bibinfo{person}{Yue Mao}, \bibinfo{person}{Yi Shen},
  \bibinfo{person}{Chao Yu}, {and} \bibinfo{person}{Longjun Cai}.}
  \bibinfo{year}{2021}\natexlab{}.
\newblock \showarticletitle{A joint training dual-mrc framework for aspect
  based sentiment analysis}. In \bibinfo{booktitle}{\emph{AAAI}},
  Vol.~\bibinfo{volume}{35}. \bibinfo{pages}{13543--13551}.
\newblock


\bibitem[\protect\citeauthoryear{Mikolov, Sutskever, Chen, Corrado, and
  Dean}{Mikolov et~al\mbox{.}}{2013}]%
        {word2vec}
\bibfield{author}{\bibinfo{person}{Tomas Mikolov}, \bibinfo{person}{Ilya
  Sutskever}, \bibinfo{person}{Kai Chen}, \bibinfo{person}{Greg~S Corrado},
  {and} \bibinfo{person}{Jeff Dean}.} \bibinfo{year}{2013}\natexlab{}.
\newblock \showarticletitle{Distributed representations of words and phrases
  and their compositionality}.
\newblock \bibinfo{journal}{\emph{NIPS}}  \bibinfo{volume}{26}
  (\bibinfo{year}{2013}).
\newblock


\bibitem[\protect\citeauthoryear{Rendle, Freudenthaler, and
  Schmidt-Thieme}{Rendle et~al\mbox{.}}{2010}]%
        {FPMC}
\bibfield{author}{\bibinfo{person}{Steffen Rendle}, \bibinfo{person}{Christoph
  Freudenthaler}, {and} \bibinfo{person}{Lars Schmidt-Thieme}.}
  \bibinfo{year}{2010}\natexlab{}.
\newblock \showarticletitle{Factorizing personalized markov chains for
  next-basket recommendation}. In \bibinfo{booktitle}{\emph{WWW}}.
  \bibinfo{pages}{811--820}.
\newblock


\bibitem[\protect\citeauthoryear{Schroff, Kalenichenko, and Philbin}{Schroff
  et~al\mbox{.}}{2015}]%
        {tripletLoss}
\bibfield{author}{\bibinfo{person}{Florian Schroff}, \bibinfo{person}{Dmitry
  Kalenichenko}, {and} \bibinfo{person}{James Philbin}.}
  \bibinfo{year}{2015}\natexlab{}.
\newblock \showarticletitle{Facenet: A unified embedding for face recognition
  and clustering}. In \bibinfo{booktitle}{\emph{CVPR}}.
  \bibinfo{pages}{815--823}.
\newblock


\bibitem[\protect\citeauthoryear{Sun, Qian, Chen, Liang, Nguyen, and Yin}{Sun
  et~al\mbox{.}}{2020}]%
        {LSTPM}
\bibfield{author}{\bibinfo{person}{Ke Sun}, \bibinfo{person}{Tieyun Qian},
  \bibinfo{person}{Tong Chen}, \bibinfo{person}{Yile Liang},
  \bibinfo{person}{Quoc Viet~Hung Nguyen}, {and} \bibinfo{person}{Hongzhi
  Yin}.} \bibinfo{year}{2020}\natexlab{}.
\newblock \showarticletitle{Where to Go Next: Modeling Long-and Short-Term User
  Preferences for Point-of-Interest Recommendation}. In
  \bibinfo{booktitle}{\emph{AAAI}}. \bibinfo{pages}{214--221}.
\newblock


\bibitem[\protect\citeauthoryear{Ungar}{Ungar}{2005}]%
        {ungar2005analytic}
\bibfield{author}{\bibinfo{person}{Abraham~A Ungar}.}
  \bibinfo{year}{2005}\natexlab{}.
\newblock \bibinfo{booktitle}{\emph{Analytic hyperbolic geometry: Mathematical
  foundations and applications}}.
\newblock \bibinfo{publisher}{World Scientific}.
\newblock


\bibitem[\protect\citeauthoryear{Yang, Fankhauser, Rosso, and
  Cudre-Mauroux}{Yang et~al\mbox{.}}{2020}]%
        {flashback}
\bibfield{author}{\bibinfo{person}{Dingqi Yang}, \bibinfo{person}{Benjamin
  Fankhauser}, \bibinfo{person}{Paolo Rosso}, {and} \bibinfo{person}{Philippe
  Cudre-Mauroux}.} \bibinfo{year}{2020}\natexlab{}.
\newblock \showarticletitle{Location Prediction over Sparse User Mobility
  Traces Using RNNs: Flashback in Hidden States}. In
  \bibinfo{booktitle}{\emph{IJCAI}}. \bibinfo{pages}{2184--2190}.
\newblock


\bibitem[\protect\citeauthoryear{{Yang}, {Zhang}, {Chen}, and {Quc}}{{Yang}
  et~al\mbox{.}}{2015}]%
        {globalScaleData}
\bibfield{author}{\bibinfo{person}{Dingqi {Yang}}, \bibinfo{person}{Daqing
  {Zhang}}, \bibinfo{person}{Longbiao {Chen}}, {and} \bibinfo{person}{Bingqing
  {Quc}}.} \bibinfo{year}{2015}\natexlab{}.
\newblock \showarticletitle{NationTelescope: Monitoring and Visualizing
  Large-Scale Collective Behavior in LBSNs}.
\newblock \bibinfo{journal}{\emph{Journal of Network and Computer
  Applications}} \bibinfo{volume}{0}, \bibinfo{number}{0}
  (\bibinfo{year}{2015}), \bibinfo{pages}{1--16}.
\newblock


\bibitem[\protect\citeauthoryear{Yu, Cui, Guo, Lu, Li, and Lu}{Yu
  et~al\mbox{.}}{2020}]%
        {categoryAware}
\bibfield{author}{\bibinfo{person}{Fuqiang Yu}, \bibinfo{person}{Lizhen Cui},
  \bibinfo{person}{Wei Guo}, \bibinfo{person}{Xudong Lu},
  \bibinfo{person}{Qingzhong Li}, {and} \bibinfo{person}{Hua Lu}.}
  \bibinfo{year}{2020}\natexlab{}.
\newblock \showarticletitle{A Category-Aware Deep Model for Successive POI
  Recommendation on Sparse Check-in Data}. In \bibinfo{booktitle}{\emph{WWW}}.
  \bibinfo{pages}{1264--1274}.
\newblock


\bibitem[\protect\citeauthoryear{Zang, Han, Li, Wan, and Wang}{Zang
  et~al\mbox{.}}{2021}]%
        {zang2021cha}
\bibfield{author}{\bibinfo{person}{Hongyu Zang}, \bibinfo{person}{Dongcheng
  Han}, \bibinfo{person}{Xin Li}, \bibinfo{person}{Zhifeng Wan}, {and}
  \bibinfo{person}{Mingzhong Wang}.} \bibinfo{year}{2021}\natexlab{}.
\newblock \showarticletitle{CHA: Categorical Hierarchy-based Attention for Next
  POI Recommendation}.
\newblock \bibinfo{journal}{\emph{TOIS}} \bibinfo{volume}{40},
  \bibinfo{number}{1} (\bibinfo{year}{2021}), \bibinfo{pages}{1--22}.
\newblock


\bibitem[\protect\citeauthoryear{Zhang, Sun, Zhang, Kloeden, and Klanner}{Zhang
  et~al\mbox{.}}{2020}]%
        {hct}
\bibfield{author}{\bibinfo{person}{Lu Zhang}, \bibinfo{person}{Zhu Sun},
  \bibinfo{person}{Jie Zhang}, \bibinfo{person}{Horst Kloeden}, {and}
  \bibinfo{person}{Felix Klanner}.} \bibinfo{year}{2020}\natexlab{}.
\newblock \showarticletitle{Modeling hierarchical category transition for next
  POI recommendation with uncertain check-ins}.
\newblock \bibinfo{journal}{\emph{Information Sciences}}  \bibinfo{volume}{515}
  (\bibinfo{year}{2020}), \bibinfo{pages}{169--190}.
\newblock


\bibitem[\protect\citeauthoryear{Zhang, Sun, Zhang, Lei, Li, Wu, Kloeden, and
  Klanner}{Zhang et~al\mbox{.}}{2021a}]%
        {zhang2021interactive}
\bibfield{author}{\bibinfo{person}{Lu Zhang}, \bibinfo{person}{Zhu Sun},
  \bibinfo{person}{Jie Zhang}, \bibinfo{person}{Yu Lei}, \bibinfo{person}{Chen
  Li}, \bibinfo{person}{Ziqing Wu}, \bibinfo{person}{Horst Kloeden}, {and}
  \bibinfo{person}{Felix Klanner}.} \bibinfo{year}{2021}\natexlab{a}.
\newblock \showarticletitle{An interactive multi-task learning framework for
  next POI recommendation with uncertain check-ins}. In
  \bibinfo{booktitle}{\emph{IJCAI}}. \bibinfo{pages}{3551--3557}.
\newblock


\bibitem[\protect\citeauthoryear{Zhang, Yang, Abbas, Deng, Li, and Zhang}{Zhang
  et~al\mbox{.}}{2021b}]%
        {SNPR}
\bibfield{author}{\bibinfo{person}{Mingwei Zhang}, \bibinfo{person}{Yang Yang},
  \bibinfo{person}{Rizwan Abbas}, \bibinfo{person}{Ke Deng},
  \bibinfo{person}{Jianxin Li}, {and} \bibinfo{person}{Bin Zhang}.}
  \bibinfo{year}{2021}\natexlab{b}.
\newblock \showarticletitle{SNPR: A Serendipity-Oriented Next POI
  Recommendation Model}. In \bibinfo{booktitle}{\emph{CIKM}}.
  \bibinfo{pages}{2568--2577}.
\newblock


\bibitem[\protect\citeauthoryear{{Zhao}, {Zhu}, {Liu}, {Xu}, {Li}, {Zhuang},
  {Sheng}, and {Zhou}}{{Zhao} et~al\mbox{.}}{2019}]%
        {stgcn}
\bibfield{author}{\bibinfo{person}{Pengpeng {Zhao}}, \bibinfo{person}{Haifeng
  {Zhu}}, \bibinfo{person}{Yanchi {Liu}}, \bibinfo{person}{Jiajie {Xu}},
  \bibinfo{person}{Zhixu {Li}}, \bibinfo{person}{Fuzhen {Zhuang}},
  \bibinfo{person}{Victor~S. {Sheng}}, {and} \bibinfo{person}{Xiaofang
  {Zhou}}.} \bibinfo{year}{2019}\natexlab{}.
\newblock \showarticletitle{Where to Go Next: A Spatio-Temporal Gated Network
  for Next POI Recommendation}. In \bibinfo{booktitle}{\emph{AAAI}}.
  \bibinfo{pages}{5877--5884}.
\newblock


\bibitem[\protect\citeauthoryear{Zhao, Liu, Zha, and Liu}{Zhao
  et~al\mbox{.}}{2021}]%
        {zhao2021hierarchical}
\bibfield{author}{\bibinfo{person}{Yanyan Zhao}, \bibinfo{person}{Jingyi Liu},
  \bibinfo{person}{Daren Zha}, {and} \bibinfo{person}{Kai Liu}.}
  \bibinfo{year}{2021}\natexlab{}.
\newblock \showarticletitle{Hierarchical and Multi-Resolution Preference
  Modeling for Next POI Recommendation}. In \bibinfo{booktitle}{\emph{IJCNN}}.
  IEEE, \bibinfo{pages}{1--8}.
\newblock


\end{thebibliography}

\end{document}